# Cognitive dissonance or p-prims? Towards identifying the best way to overcome misconceptions in physics

Dr Panos Athanasopoulos, Prestatyn High School, UK[1]

**Abstract**

In this classroom-based action-research project, I compared the following two approaches to check their effectiveness in helping students overcome physics misconceptions: Inducing cognitive dissonance or gradually building on students' previous knowledge activating the relevant phenomenological primitives (p-prims). This took place over a two-lesson sequence (each an hour long) using year 8 (12 years old) and year 9 (13 years old) top set students (N=87 in total), in the context of Newton's first law. Results were better for both year groups when inducing cognitive dissonance, which seems to be more effective not only with surface-level learning, but deep-learning as well.

**Introduction and rationale**

David Ausubel has famously described prior domain knowledge as the most important determinant of a student's learning success: "Ascertain what they know and teach them accordingly" (Cook & Ausubel, 1970). However, a recent meta-analysis (Simonsmeier et al, 2022) suggests that this is not the end of the story. As expected, students who know most at the beginning will tend to know most after instruction as well. Nonetheless, students' prior knowledge cannot (on average) explain the variance in learning gains, i.e. students with the highest prior knowledge do not (on average) make the most progress.

This raises interesting questions and (Brod, 2021) describes a framework within which this could be investigated further in future meta-analyses. In this paper, I will attempt to approach the topic within the context of physics teaching, where there are concrete mechanisms to explain why prior knowledge might hinder learning and significant work has been done on the topic of misconceptions and conceptual change.

**Literature review**

Physics is quite unique among the subjects because students do not arrive in the lessons with a "blank slate". They already have experience of the physical world and have developed their own conceptions about how things work. These conceptions can often clash with the conceptions of experts in the field. The terms misconceptions, preconceptions, alternative conceptions or naïve conceptions are commonly used in the literature to describe this phenomenon. This creates some tension because it is well established that activating prior knowledge is an important component of instruction (Rosenshine, 2012). However, in science, simply activating existing knowledge is detrimental if the existing knowledge is wrong (Poehnl & Bogner, 2013). Learning in physics is often about modifying existing conceptions rather than developing a schema from scratch.

---

[1] pathanasopoulos@prestatynhigh.co.uk

But how important is it for the teacher to be aware of the misconceptions students hold? After all, it is conceivable that by just teaching the correct idea the students will eventually pick it up and overcome their misconceptions. Unfortunately, it looks like this might not be the case. (Sadler et al, 2013) ran a large-scale experiment (9,556 students of 181 middle school physical science teachers in the USA) testing both teachers and students using appropriately designed multiple choice questions. They found that "For items that had a very popular wrong answer, the teachers who could identify this misconception had larger classroom gains, much larger than if the teachers knew only the correct answer."

Some caution is in order here, because this correlation is not proof of causation. One could imagine for example that there is an intermediate variable such as "general teaching skill" or "years of teaching" which is the key contributor to the teachers correctly identifying the misconceptions and student gains as well. That being said, the big sample of participating students indicates that the correlation discovered is strong and this should be taken into account.

Further supportive evidence comes from (Muller, Sharma, & Reimann, 2008) who claim that the teacher needs to not only be aware of possible alternative conceptions the students might hold but to actively address them in order to maximize student gain. This phenomenon appears to be rather robust and seems to persist even when an idea is taught through a video rather than face-to-face (Muller et al., 2008). However, when and how to address the misconceptions seems to be under-investigated in certain respects. There are, broadly speaking, two alternative approaches that could be followed and the purpose of this investigation is to compare them.

The first approach is the *elicit, confront, resolve* (McDermott & Shaffer, 2002) method or other variations based on this theme such as the *Elicit-Confront-Identify-Resolve-Reinforce* (Wenning & Vieyra, 2020). Here the teacher begins instruction by asking students questions that they know will bring up possible misconceptions. The students are then asked to commit to their answers (e.g. by writing them down) and the teacher proceeds to demonstrate through an experiment or otherwise why those conceptions are clearly wrong. The students are then left in a state of cognitive dissonance and are allegedly open to learning better and updating their mental-schema of the topic.

It might seem counter-intuitive that cognitive dissonance help students learn well, but cognitive dissonance theory has actually been around for many decades in the field of psychology (see for example (Harmon-Jones & Mills, 2019) for an introduction). One possible mechanism of why it works could be that the existence of dissonance, being psychologically uncomfortable, motivates the person to work towards reducing the dissonance. Alternatively, maybe the effect stimulates curiosity and a love of learning something new. No matter what the underlying mechanism is, the approach is well researched in psychology as well as in physics (Wenning & Vieyra, 2020) and it consistently works. What we are trying to investigate here is if the alternative is even better, or not.

The alternative approach is heavily based on the concept of phenomenological-*primitives* (or p-prims for short) (diSessa, 1993; Hammer, 1996; Smith, DiSessa, & Roschelle, 1994). A p-prim is a pattern of thought that is applied across a range of contexts. For example, "Increasing the cause will increase the effect.", "Reducing resistance will increase the effect.", "Removing the cause will make the effect die away", etc are all examples of p-prims. P-prims are more fundamental, more abstract cognitive structures that the rest of our physics knowledge is based on, in a way somewhat similar to how mathematics is based on axioms.

The idea of p-prims brings about a fundamental change to how a teacher should approach the lesson. (Hammer, 1996) discusses the example of how, in a popular demonstration of misconceptions, students were asked to explain why it is hotter in the summer than in the winter (Sadler, Schneps, & Woll, 1989). Many responded that this is because the earth is closer to the sun, rather than giving the correct answer that it is because of the tilt of the earth relative to the sun. It could simply be that students store the wrong information in their long-term memory and we need them to replace this. However, the p-prims perspective does not attribute a knowledge structure concerning the closeness of the sun and the earth. It claims instead that it could be that the students are erroneously activating the p-prim "Closer means stronger." The p-prim itself is not wrong: Candles are hotter and brighter the closer you get to them; music is louder the closer you are to the speaker; the smell of garlic is more intense the closer you bring it to your nose. It is the application of the p-prim to this particular phenomenon that is wrong and the teacher should guide the students to associate the correct p-prim ("Tilting away reduces intensity.") with this case.

To achieve this, the intention is to teach and practise the correct idea without forcing the students to commit or experience the cognitive dissonance. This is because we would like to avoid activating the wrong p-prim, which might naturally happen if you force them to commit to an answer. We are instead trying to go "under the radar" of possible alternative conceptions they might hold and practise examples of the correct p-prim to prepare the students for what is coming. After the students have practised, they become more familiar with the idea/schema in the version held by the expert teacher and more used to applying the correct (for this particular case) p-prim. It is only then, that the teacher brings into their attention how the newly developed schema might be in contradiction to certain naïve conceptions they might have held up to now or that many other people hold. The teacher also explains why the new schema is superior and the evidence backing it up further.

There are further theoretical considerations as to why the second approach might be successful. The first one is cognitive load theory (Sweller, van Merriënboer, & Paas, 2019). By getting the students familiar with the new idea in a somewhat isolated context (before connections are made to aspects of life in which the students might hold alternative conceptions), we would expect a reduced cognitive load during the actual "confrontation" which will allows students to see the bigger picture and will free up resources for the comparison, rather than leaving the students overwhelmed with trying to understand a new concept as well as modify their existing alternative conceptions.

That being said, both approaches are reasonable and have their advantages. It can be argued for example that the first one generates a lot of interest in the topic and motivates the students to find out why their ideas are wrong. On the other hand, in the second approach the students have already learned and practised the new idea by the time they are exposed to the contradiction. Therefore, they should arguably experience a reduced cognitive load, which should make it easier for them to follow the argument and process the information. Furthermore, both approaches have been shown to improve student outcomes. It is the goal of this project to trial and compare these two methods in the context of year 8 and year 9 physics and in the topic of motion and forces. For the remaining of this paper and to facilitate the data analysis we will refer to these two approaches as "cognitive dissonance" and "p-prim" approach, with the exact implementation for each described in the next section.

## Methodology

The two top sets from year 8 (8p1 and 8p2) and year 9 (9p1 and 9p2) were selected for this investigation. The number of students involved is given in Table 1. Newton's first law was selected as the topic to be taught and each group was taught two one-hour lessons on forces and motion. The first lesson was identical for all groups and was intended to establish the notion of forces, the different types of forces and the basic vocabulary that will be used in the second lesson. There were two main learning outcomes for the first lesson. The first one was that students should be able to identify a variety of forces on objects including forces from hands, springs, magnets, normal contact force and weight. The second was that students understand that forces appear when two objects interact.

|  | Pupils in set | Pupils that attended both lessons and took the required tests | Type of instruction |
|---|---|---|---|
| 8p1 | 31 | 22 | Cognitive dissonance |
| 8p2 | 33 | 18 | p-prim |
| 9p1 | 32 | 24 | p-prim |
| 9p2 | 34 | 23 | Cognitive dissonance |

Table 1. Numbers of students involved and type of instruction for each set.

The second lesson began with questioning about the ideas from lesson 1. For the two groups receiving the cognitive dissonance instruction, the students were then challenged to identify which force keeps a paper plane flying through the room. Students were allowed to brain storm and I explained why each of the possible answers was wrong. When no more guesses were offered, I explained that objects do not need a force acting on them in order to be moving and the rest of the lesson was used to elaborate the point further using the phet motion simulation and more demos.

The second lesson for the p-prim groups also began with questioning about the ideas from lesson 1 but then moved straight into the phet motion simulation and the demos. After the students were familiar with the phenomena, I teased out the idea that objects do not need a force acting on them in order to be moving and checked for understanding. Finally, I brought to their attention that this might be in contradiction to what we might expect from every experience and the students readily offered that this is because of friction and air-resistance.

Every effort was made to keep the instruction between the different groups as identical as possible, with the only difference being whether cognitive dissonance was induced at the beginning of the second lesson or not.

## Ethical considerations

Permission was granted by the headteacher for this research to take place. Normal science lessons were used and the topic taught is part of the standard curriculum.

## Data collection process

The test given to the students can be found online here (link). It is a modified version of a diagnostic assessment developed through the Evidence based Practice in Science Education framework at the University of York (link to original assessment here). The same test was used as a pre-test (where applicable, see below) and a post-test. Some questions were removed from the original to reflect the time constraints of the lessons and the content taught and some were slightly modified to align better with the way I was teaching the content. Nevertheless, in my professional judgement the validity of this assessment is high and I would urge the interested reader to inspect the version used and satisfy themselves that it is indeed testing knowledge of Newton's 1$^{st}$ law. As a matter of fact, some questions dubbed "high discriminators" appear to be testing deep, rather than the surface knowledge. We will come back to these later.

Because of the nature of this research, it is conceivable that giving the test as a pre-test to the students and forcing them to commit their answers on paper might induce cognitive conflict either during the test itself or the early parts of instruction that followed it. This could potentially invalidate our results, because our goal is to avoid cognitive conflict and cueing the wrong p-prim in the groups receiving the p-prim instruction. On the other hand, having concrete data that show progress in terms of both a pre-test and a post-test can strengthen the validity of our results.

To balance the two opposing forces and to mitigate the risk of invalidating the results I decided to proceed as follows: Year 8 received no pre-test. Since they have had no exposure to the topic before, I felt safe to assume that they will have the usual misconceptions and that they are all having a very similar starting point. This is an assumption I am happy to make, because in all my years as a physics teacher, I have never met a student who intuitively discovered and understood Newton's first law without having being taught it. This also seems to be the case for hundreds of students across several countries (Bliss, Ogborn, & Whitelock, 1989; Watts, 1983). (Driver et al, 2014) offer a good review of the literature on this topic. For this year group, we will be comparing the **scores** in the post-test given at the end of the second lesson to see which group performed better.

Year 9 received a pre-test at the very beginning of the first lesson and a post-test at the end of the second lesson. This year group has had some exposure to forces 3-4 months before this research project. Though Newton's first law was not meant to be part of it, it is conceivable that some of the teachers covered at least some aspects of it. It was therefore safer to check the starting point of the participating students. We will use these data to compare **progress** between the two groups.

**Further limitations and potential biases**

Students in our school are set by ability and performance in previous exams, so we would naturally expect that students in 8p1 would most likely outperform those in 8p2 in any test and likewise for 9p1 and 9p2. I tried to account for this by offering different types of instruction in 8p1 and 9p1 as described in Table 1 (and similarly for 8p2 and 9p2).

I had also discussed aspects of Newton's 1$^{st}$ law with the year 9 groups that I normally teach during the year. To minimise any initial exposure to the subject, I ran this project with groups that I do not normally teach and with students that I had not interacted with before this investigation.

To minimise my own bias about which method I thought would yield superior results (the p-prim instruction), I marked the pre-tests as soon as the students had taken them but did not mark any of the final tests until instruction of all groups had finished.

The final restriction is the number of participating students. This investigation ran at the end of the academic year 2021-22, at a time when some students were out of lessons for a variety of activities, resulting in smaller numbers than I initially expected (see Table 1). That being said, all my results are self-consistent among the groups and provide some interesting insights.

## High discriminators

After marking the pre-tests, it became obvious that certain questions were easily answered by the vast majority of students even with no previous exposure to Newton's 1st law. Such a question can be seen in Figure 1.

On the other hand, 3 out of the 16 questions on the test were almost universally answered wrong in the pre-test. My interpretation of this is that these questions were testing deeper, rather than surface, understanding of Newton's 1st law. I will be referring to these 3 questions as "high discriminators". An example of a high discriminator can be seen in Figure 2.

I performed a separate analysis based on answers to the entire test and on high discriminators only and the results can be found in the next section.

(a) **The car is slowing down.** Which of the following best describes the size of these two forces?

Tick ONE box (✓)

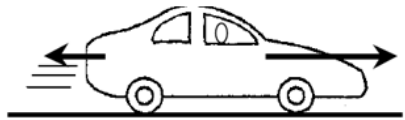
☐ The driving force is **bigger** than the counter force.

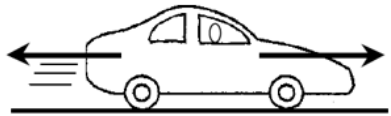
☐ The driving force is **exactly the same size** as the counter force.

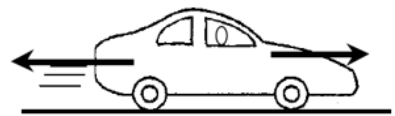
☐ The driving force is **smaller than** the counter force.

Figure 1. Example of a question that was easily answered by the vast majority of students even with no previous exposure to Newton's 1st law.

Andy kicks a football across a level pitch. It rolls from A to C where it stops.

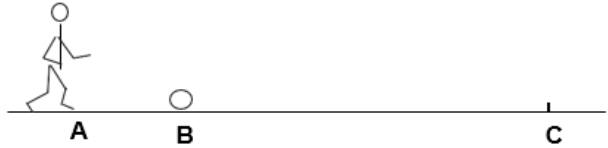

(b) Which of the following diagrams best shows the **horizontal** forces on the ball at point **B**?
Tick ONE box (✓)

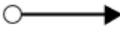

Figure 2. Example of a high discriminator, i.e. a question that was testing for deeper understanding.

### Key findings

### Year 8

The key data for year 8 are presented in graph 1 and table 2. The 8p1 group receiving the cognitive dissonance instruction has outperformed the 8p2 group. Treating 8p2 as the control group, we can calculate the effect size of the cognitive dissonance instruction to be 0.92 with the 95% confidence interval being (0.23-1.61). (For effect size calculations I follow the conventions of (Coe, 2002) and the pooled standard deviation was used.) This seems to imply that inducing cognitive dissonance had a medium to high effect in the learning of the students. When looking at the high discriminators only, progress seems to be negligible in either form of instruction. The effect size of cognitive dissonance is calculated to be -0.14 with the 95% confidence interval being (-0.75, 0.47). This seems to indicate that two lessons are not sufficient for students to gain a deep understanding of Newton's 1st law and overcome those deep-rooted misconceptions.

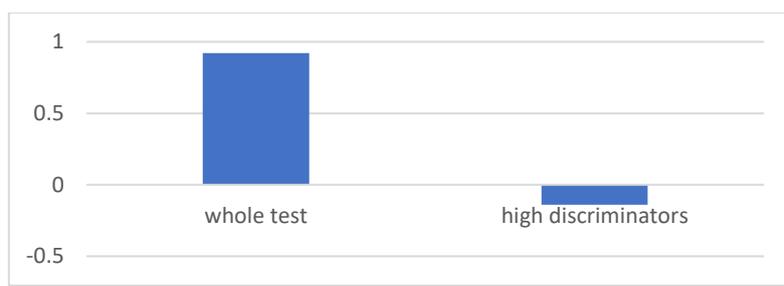

Graph 1. Year 8 effect size of cognitive dissonance method taking p-prim method as the control group.

|  | 8p1 (cognitive dissonance) | 8p2 (p-prim) |
|---|---|---|
| Mean posttest score | 65% | 48% |
| Standard deviation | 8% | 25% |
| Mean posttest score on high discriminators | 9% | 12% |
| Standard deviation | 15% | 19% |

Table 2. Year 8 results.

**Year 9**

The key data for year 9 can be found in graph 2 and tables 3 and 4. The group receiving the cognitive dissonance instruction has also outperformed the other group. Even though the gain for both groups might look identical, the much smaller standard deviation in 9p2 means that this is more likely a real effect than a statistical fluctuation. This becomes much more apparent when calculating the effect sizes, which can be found in table 3. Once again, inducing cognitive dissonance appears to be superior and has a medium to high effect in the learning of the students. For the 9p2 group, cognitive dissonance also seems to have resulted to some deep learning, as demonstrated by the gain in the high discriminators, which is an interesting result by itself because this was not the case for any other group. Finally, we note that in physics education literature the measure most commonly used is the average normalized gain ⟨g⟩, which is defined as the ratio of the actual average gain (%⟨post⟩−%⟨pre⟩) to the maximum possible average gain (100−%⟨pre⟩). This number can be thought of as the percentage of marks the students got correct after instruction out of all the marks that they initially got wrong. For completeness, we include this number in the table as well.

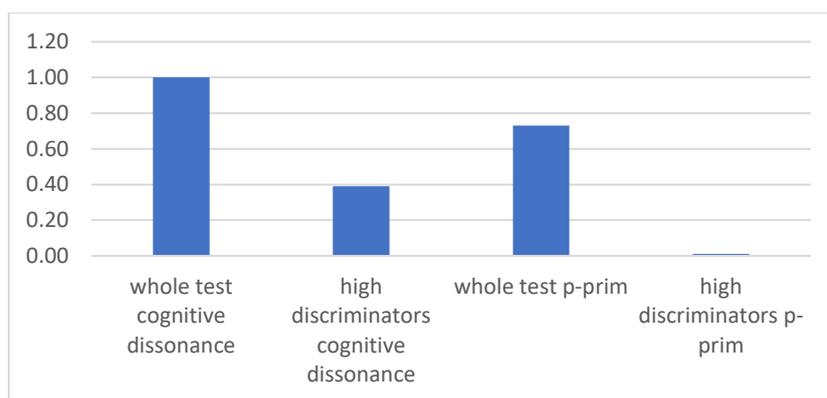

Graph 2. Year 9 effect size of the two different instruction types using pre-test and post-test scores.

|  | 9p1 (p-prim) | 9p2 (cognitive dissonance) |
|---|---|---|
| Effect size (whole test) | 0.73 | 1.00 |
| 95% confidence interval | (0.11-1.34) | (0.36-1.65) |
| Effect size (high discriminators) | 0.00 | 0.39 |
| 95% confidence interval | (-0.57, 0.57) | (-0.21, 1.00) |

Table 3. Year 9 effect size and confidence interval for the different instruction methods.

|  | 9p1 (p-prim) | 9p2 (cognitive dissonance) |
|---|---|---|
| Mean pretest score | 50% | 43% |
| Standard deviation | 23% | 14% |
| Mean posttest score | 65% | 58% |
| Standard deviation | 14% | 16% |
| Gain (whole test) | 14% | 15% |
| average normalized gain ⟨g⟩ | 0.28 | 0.26 |
| Mean pretest score (high discriminators) | 13% | 6% |
| Standard deviation | 21% | 16% |
| Mean posttest score (high discriminators) | 13% | 16% |
| Standard deviation | 27% | 32% |
| Gain (high discriminators) | 0% | 10% |
| average normalized gain ⟨g⟩ (high discriminators) | 0 | 0.10 |

Table 4. Year 9 results (numbers rounded to two significant figures).

**Conclusion and recommendations**

Contrary to what I expected before starting this inquiry, inducing cognitive dissonance at the beginning of a learning sequence seems to be superior to explaining common misconceptions at the end. This is somewhat counter-intuitive because it appears to be going against the ideas of modern cognitive load theory as described for example in (Sweller et al., 2019).

In my opinion, the contradiction is only superficial and can probably be explained by realising that cognitive dissonance could be inducing germane cognitive load which is required for learning. Furthermore, It is likely that it creates a "desirable difficulty" (Bjork & Bjork, 2011) which is connected to learning better.

The recommendation would therefore be to try and use the cognitive dissonance approach when appropriate. Topics where it is expected the students will have deep-rooted misconceptions such as Newton's 1st law lent themselves well to this style of teaching. There will of course be other topics where students are simply unfamiliar with the context and there are no strong misconceptions and the teacher can use their professional judgement as to the best way of teaching those.

However, when using the cognitive dissonance approach, some caution is necessary. This project shows that cognitive dissonance works when the students are well acquainted with all the terms and ideas that are discussed and the teacher should spend sufficient time to ensure that this is the case. I would speculate that inducing cognitive dissonance on a topic the students are unfamiliar with, using terms they have not heard of before will not result in much learning no matter how engaging the experiment or the activity is. Given of course the fact that my prediction on this project was wrong, this is something that should be investigated properly.

The final argument in favour of inducing cognitive dissonance at the beginning was that it made the lessons more interesting for everyone involved. This is a subjective judgement on my behalf based on a small sample of lessons, but the lessons I tried it in definitely had a different vibe and spark in them and the engagement of the students appeared increased. This is an important positive takeaway because many of us became physicists exactly because of these sparks of interest and the sense of wonder as to how our world works that they generate. It is great to see that teaching this way also appears to be best for learning.